\documentclass{jpsj2}

\usepackage{subfigure}

\title{
Crossover from Positive to Negative Interlayer Magnetoresistance
in Multilayer Massless Dirac Fermion System
with Non-Vertical Interlayer Tunneling
}

\author{Takao \textsc{Morinari}\thanks{
morinari@yukawa.kyoto-u.ac.jp
}
and Takami \textsc{Tohyama}
}

\inst{
Yukawa Institute for Theoretical Physics, Kyoto University, Kyoto 606-8502, Japan
}

\abst{
We present a theoretical description of the interlayer magnetoresistance 
in the layered Dirac fermion system with the application to
the organic conductor $\alpha$-(BEDT-TTF)$_2$I$_3$ under pressure.
Assuming a non-vertical interlayer tunneling and including higher Landau level effects
we calculate the interlayer conductivity using the Kubo formula.
We propose a physical picture of the experimentally observed crossover from the negative 
interlayer magnetoresistance, where the Dirac fermion zero-energy Landau level 
plays a central role, to the positive interlayer magnetoresistance
that is the consequence of the Landau level mixing effect 
upon non-vertical interlayer hopping.
The crossover magnetic field depends on the Landau level broadening factor
and can be used to determine the Dirac fermion Landau level energy spectrum.
}

\kword{
$\alpha$-(BEDT-TTF)$_2$I$_3$, Dirac fermions, interlayer magnetoresistance}

\newcommand{\refeq}[1]{(\ref{#1})}                                  

\newcommand{\be}{\begin{equation}}
\newcommand{\ee}{\end{equation}}
\newcommand{\bea}{\begin{eqnarray}}
\newcommand{\eea}{\end{eqnarray}}

\begin{document}
\maketitle

\section{Introduction}
In condensed matter systems, it is possible
to realize a massless Dirac fermion like excitation
spectrum of electrons with the speed of light 
being replaced by a Fermi velocity.
For example the electron energy dispersion described by the tight-binding model 
with nearest neighbor hopping on the honeycomb lattice
are well approximated by a massless Dirac fermion spectrum
at the high-symmetry points at the corners of the first Brillouin zone.\cite{Wallace47}
Since the discovery of unconventional integer quantum Hall effect 
in graphene,\cite{Novoselov2005,Zhang2005}
which is a two-dimensional carbon material with a honeycomb lattice,
Dirac fermions realized in condensed matter systems
have attracted much attention.

In conventional two-dimensional electron systems at low temperatures 
and strong perpendicular magnetic fields,
the Hall conductivity $\sigma_{xy}$ exhibits plateaus at $n e^2/h$ 
with $n$ the number of filled Landau levels.\cite{Prange87}
By contrast in two-dimensional Dirac fermion systems
$\sigma_{xy}$ exhibits plateaus at $\sigma_{xy} = \pm (|n| + 1/2) e^2/h$ with
$n$ the Landau level index.\cite{Novoselov2005,Zhang2005}
(Here we do not include the degeneracy factor due to spin and valley.)
In the Dirac fermion system the energies of the Landau level are described by
\be
E_n = {\rm sgn}(n) C \sqrt{|n| B},
\ee
with $n=0, \pm 1, \pm 2, ...$ and $B$ being the magnetic field.
The shift of $e^2/(2h)$ in the Hall conductivity plateaus
is associated with the existence of the zero energy Landau level, $n=0$.
\cite{CastroNeto09}

Another massless Dirac fermion system is realized in 
$\alpha$-(BEDT-TTF)$_2$I$_3$ under pressure.\cite{Kobayashi04,Katayama06,Kobayashi07}
(Here BEDT-TTF denotes bis(ethylenedithio)-tetrathiafulvalence.)
This system has a layered structure,
and the conducting layers consist of BEDT-TTF molecules.
Between conducting layers there is an insulating iodine layer.
The system exhibits small inplane conductivity anisotropy while
the conductivity perpendicular to the plane is much less than
that in the plane.\cite{Bender84}
This strong anisotropy suggests that the system consists of
weakly interacting conducting layers.
In the ambient pressure, a metal-insulator transition occurs at $135$K \cite{Bender84}
associated with a charge ordering.\cite{Kino1995,Seo2000,Takano2001,Wojciechowski2003}
This metal-insulator transition is suppressed by applying pressure.\cite{Kartsovnik85,Schwenk85}
Under pressure larger than $\sim 2$GPa, the system is metallic and 
the resistivity is almost temperature independent while the Hall coefficient
shows strong temperature dependences.\cite{Tajima2000}
Kobayashi {\it et al.} calculated the energy dispersion \cite{Kobayashi04,Katayama06}
using the tight-binding model.
The transfer integrals were estimated from 
X-ray diffraction experiments.\cite{Kondo2005}
They found that the electronic band structure near the Fermi level 
is described by a tilted and anisotropic Dirac cone.
Such a Dirac cone structure was also obtained
by first principles calculations.\cite{Ishibashi2006,Kino2006}

The presence of the Dirac fermion spectrum was 
supported by observations of the negative interlayer 
magnetoresistance.
Tajima {\it et al.} reported a remarkable negative interlayer 
magnetoresistance under perpendicular magnetic field.\cite{Tajima09} 
The magnetic field dependence of this negative interlayer magnetoresistance 
is in good agreement with the formula given by Osada
based on the zero energy Landau level.\cite{Osada08}
For relatively large magnetic field, Osada's formula
is quantitatively in good agreement with the experiments.
In addition, in our previous publication \cite{Morinari09}
we reported that the interlayer magnetoresistance
can be used to verify the tilt of the Dirac cone
by extending the Osada's formula.
Meanwhile, the system exhibits a positive magnetoresistance at low magnetic field.
The origin of this positive magnetoresistance has not been clarified so far.

In this paper, we argue that the positive interlayer magnetoresistance
appears when the direction of the interlayer electron hopping
is not perpendicular to the plane.
Under the condition of such a non-vertical interlayer hopping
electrons move in the plane upon interlayer hopping.
Suppose electron hops from $j$-th layer to $j+1$-th layer.
The change of the electron coordinate in the plane
implies that the $n$-th Landau level wave function in $j$-th layer
is not orthogonal to the $n'$-th Landau level wave function 
in $j+1$-th layer ($n\neq n'$).
This inter-Landau level mixing leads to the positive magnetoresistance
when $|E_n - E_{n+1}|$ is less than the energy scale $\Gamma$
characterizing the Landau level broadening that is satisfied 
at low magnetic fields.

The rest of the paper is organized as follows.
In sec.\ref{sec_interlayer_hopping}, we describe the model
of the non-vertical interlayer hopping.
In sec.\ref{sec_result}, we calculate the interlayer 
magnetoresistance and show that there is crossover
between a positive interlayer magnetoresistance
to a negative interlayer magnetoresistance.
Finally we summarize the result in sec.\ref{sec_summary}.

\section{Model}
\label{sec_interlayer_hopping}
\begin{figure}[htbp]
  \begin{center}
  \subfigure[]{
    \includegraphics*[height=0.3 \linewidth,keepaspectratio]{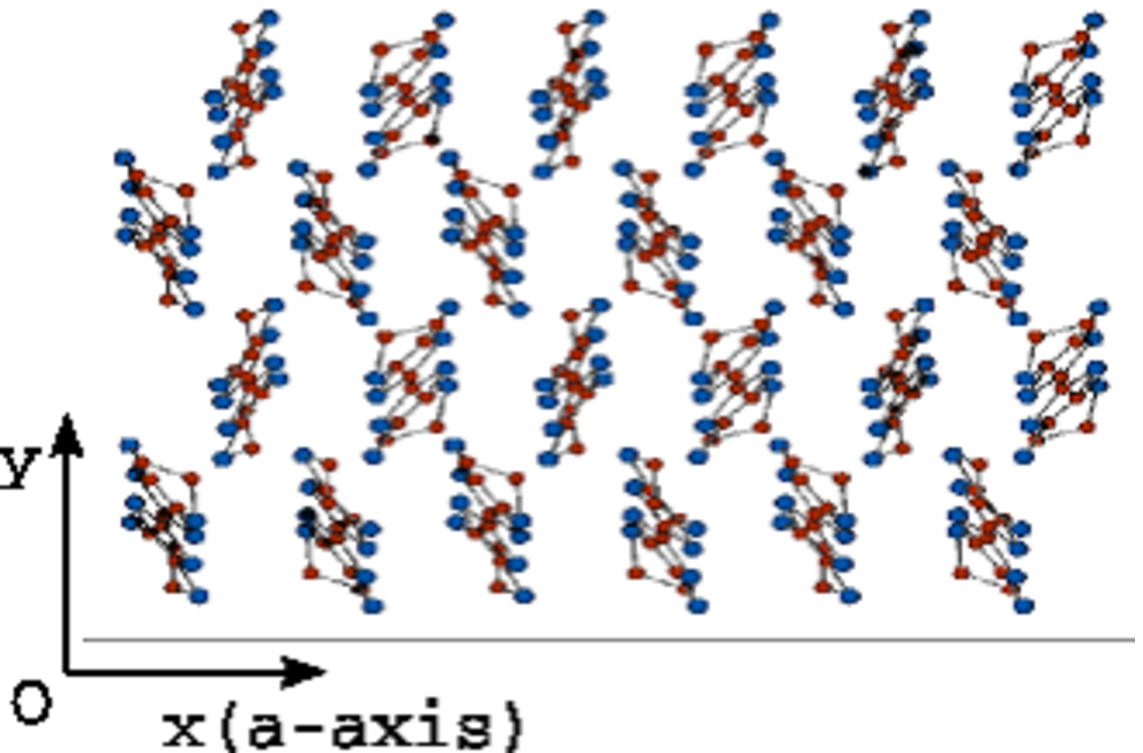}
  }
  \subfigure[]{
    \includegraphics*[height=0.3 \linewidth,keepaspectratio]{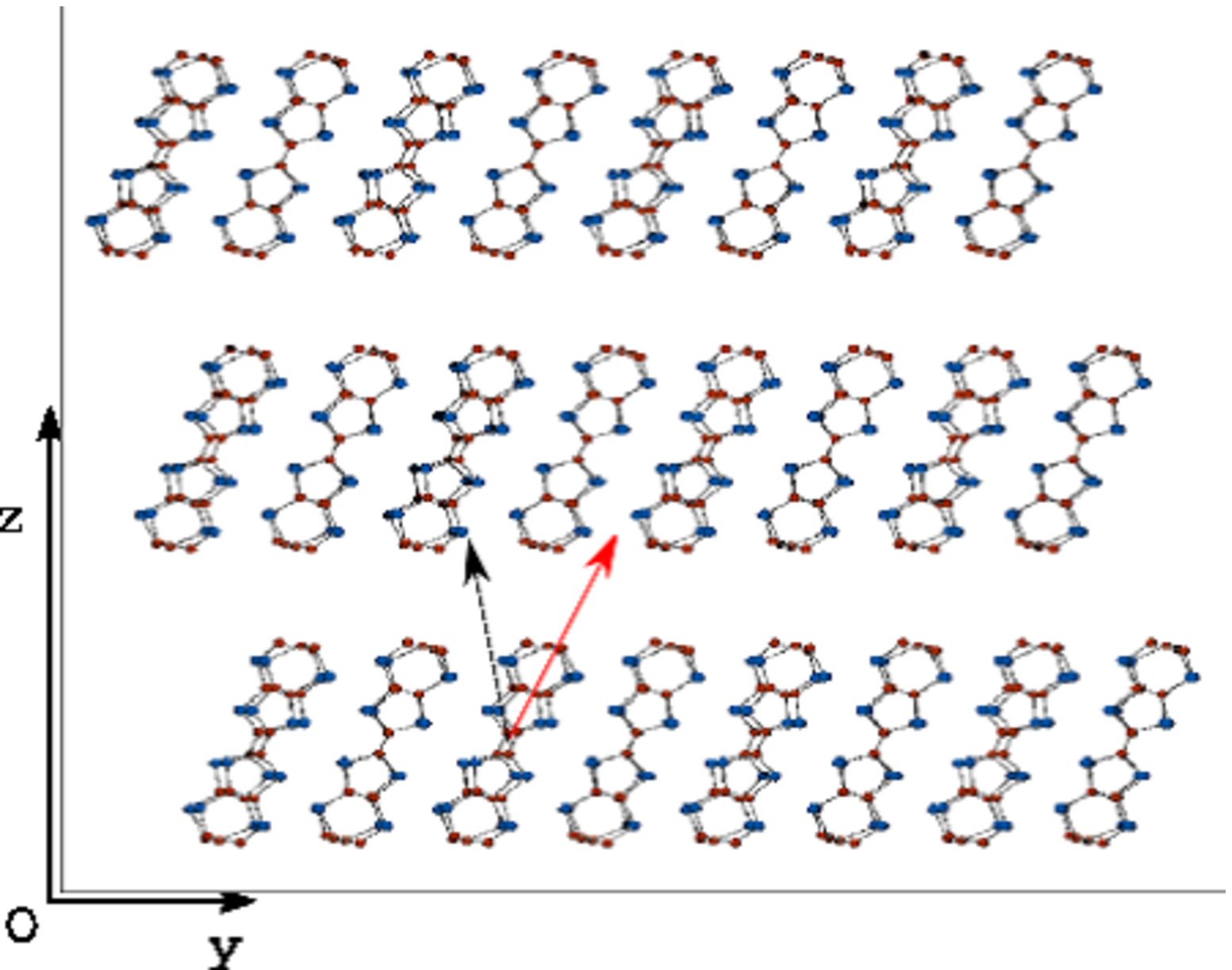}
  }
\end{center}
\caption{ 
(Color online)
(a) The xy-coordinate system defined in the conducting planes.
The $a$-axis is taken to be the $x$-axis.
The BEDT-TTF molecules form stacks parallel to the $a$-axis.
The $z$-axis is taken to be perpendicular to the plane.
(b)
The direction of the interlayer tunneling viewed from the positive $a$-axis.
The dashed arrow shows the direction of the $c$-axis.
The solid arrow shows the direction of the interlayer tunneling,
$a_b {\hat {\bf e}}_b + a_c {\hat {\bf e}}_c$,
where $a_b$ and $a_c$ are the lattice constants and 
${\hat {\bf e}}_b$ and ${\hat {\bf e}}_c$ are 
the unit vectors in the $b$ and $c$ axes, respectively.
}
\label{fig_axes}
\end{figure}

Following Ref.\citenum{Osada08}
we calculate the interlayer conductivity
by taking the interlayer transfer integral, $t_c$, 
as a small parameter.
Cartesian coordinate axes are defined in Fig.\ref{fig_axes}(a).
The direction of the interlayer tunneling is defined in Fig.\ref{fig_axes}(b).
From the crystal structure data in Ref.\citenum{Bender84}
we find that the angle between the interlayer tunneling direction
and the $z$-axis is about $28$ degree.
This angle may depend on applied pressure.
However, the precise value of the angle is not necessary.
The value of the interlayer conductivity is insensitive 
to changes of the angle as we shall see later.
The interlayer tunneling Hamiltonian is written as
\be
H_{c'}  =  - t_{c'} \int {dx} \int {dy} \sum\limits_{j,\sigma } 
{\psi _{j\sigma }^\dag  \left( {x,y} \right)\psi _{j + 1,\sigma } 
\left( {x ,y + c_y } \right)}  + h.c.
\label{eq_interlayer_tunneling}
\ee
Here the operator $\psi_{j\sigma}(x,y)^{\dagger}$ is 
the creation operator for an electron with spin $\sigma$
at point $(x,y)$ in the $j$-th plane.
The interlayer tunneling direction, which is not parallel to the $c$-axis
as shown in Fig.\ref{fig_axes}(b), is denoted by $c'$ hereafter.
There is a slight translation in the $x$-axis
in the interlayer tunneling direction defined in Fig.\ref{fig_axes}(b).
We neglect that translation so that the interlayer tunneling direction
is parameterized by $c_y$.

The interlayer current operator is found by making use of the continuity equation.
We calculate the time derivative of the density operator 
under the interlayer tunneling Hamiltonian eq.(\ref{eq_interlayer_tunneling}).
The result is equal to the divergence of the interlayer current operator
in the discretize form with the minus sign.
The effect of the magnetic field, $B$, 
is included by the Pierls substitution.
We focus on the magnetic field perpendicular to the plane.
Thus, the current operator is
\be
 {\hat J}_{c'}  = ia_z t_{c'} \sum\limits_{j,\sigma } 
{\psi _{j + 1,\sigma }^\dag \left( {x,y} \right)
\exp \left[ { ic_y \left( {{\hat k}_y  + eBx} \right)} \right]
\psi _{j,\sigma } \left( {x,y} \right)}
+ h.c.,
\label{eq_Jc}
\ee
where ${\hat k}_{\alpha} = -i\partial_{\alpha}$ ($\alpha=x,y$)
and $a_z$ is the interlayer distance.
Hereafter we employ units where $\hbar = 1$.

Within the conducting layers the dynamics of the electrons
is described by an extended Hubbard model.\cite{Kobayashi04,Katayama06,Kobayashi07}
Kobayashi {\it et al.} introduced mean fields for charge disproportionation.
Since there are four BEDT-TTF molecules in the unit cell,
four charge density mean fields were introduced.
Using the self-consistent mean field values they found the zero gap state. 
In the vicinity of the zero gap point in the Brillouin zone, called Dirac points,
the electrons are described by the effective Hamiltonian that has the form of 
a massless Dirac fermion Hamiltonian, or, more precisely, 
a tilted Weyl Hamiltonian.\cite{Kobayashi07}
There are two Dirac points in the Brillouin zone.
If there are impurity potentials with extremely short interaction range
like a lattice vacancy, there is scattering between two Dirac points.\cite{Suzuura02}
However, it is unlikely that there are lattice vacancies in the BEDT-TTF molecule planes.
We neglect intervalley scattering and we assume that
two Dirac cones are degenerate.
In this case we may focus on a single Dirac point.
The Hamiltonian describing Dirac fermions in the vicinity of the Dirac cone is
\be
{\cal H} = \sum\limits_{j,\sigma } {\int {d^2 {\bf{r}}} \,
\psi _{j\sigma }^\dag  \left( {\bf{r}} \right)\left( {\begin{array}{*{20}c}
   v \eta {\hat k}_y & {v\left( {\hat k}_x  - i{\hat k_y}  \right)}  \\
   {v\left( { {\hat k}_x  + i{\hat k}_y } \right)} & v \eta {\hat k}_y  \\
\end{array}} \right)} \,\psi _{j\sigma } \left( {\bf{r}} \right).
\label{eq_DiracH}
\ee
In order to focus on the crossover phenomenon from the negative 
to positive interlayer magnetoresistance
we have taken a simple form compared to the general form in Ref.\citenum{Kobayashi07}.
In $\alpha$-(BEDT-TTF)$_2$ I$_3$, the Dirac cone is tilted \cite{Kobayashi07}.
The parameter $\eta$ is introduced in eq.\refeq{eq_DiracH}
to describe the tilting effect.

\section{Unified Formula for Interlayer Magnetoresistance}
\label{sec_result}
Now we calculate the interlayer conductivity using the Kubo formula,
\bea
\sigma _{zz} &=&  - \frac{i}
{S}\sum\limits_{\alpha ,\beta } 
  {\frac{{f\left( {E_\alpha  } \right) - f\left( {E_\beta  } \right)}}
{{E_\alpha   - E_\beta  }}
  \frac{{\left| {\left\langle \alpha  \right| {\hat J}_{c'} \left| \beta  \right\rangle } \right|^2 }}
{{E_\alpha   - E_\beta   + i\delta }}}  \nonumber \\ 
&=& \frac{\pi }
{S}\sum\limits_{\alpha ,\beta } 
{\int_{ - \infty }^\infty  {dE} \left( { - \frac{{\partial f}}
{{\partial E}}} \right)\left| 
{\left\langle \alpha  \right|{\hat J}_{c'} \left| \beta  \right\rangle } \right|^2 
\delta \left( {E - E_\alpha  } \right)\delta \left( {E - E_\beta  } \right)}.
\label{eq_kubo_formula}
\eea
where $S$ is the area of the conducting layers,
$f=f(E)$ denotes the Fermi distribution function, 
$\delta$ is a positive infinitesimal number,
and $\alpha$ and $\beta$ represent the single-body quantum states.
The eigen-energy of the quantum state $\alpha$ is represented by $E_{\alpha}$.
The current operator is defined by eq.(\ref{eq_Jc}).
In order to include impurity scattering effects in eq.\refeq{eq_kubo_formula}, 
we replace the delta functions,
which are associated with the density of states,
with the Lorentzian function with the half value width of $\Gamma$.
We take $\Gamma$ as a constant parameter for simplicity.

In a magnetic field $B$ parallel to the $z$-axis, the electron motion in the plane
is quantized to the Landau levels.
The derivation of the Landau level wave functions is presented 
in Appendix.
If we take the Landau gauge, ${\bf A}=(0,Bx,0)$, the Hamiltonian 
does not depend on $y$.
So the wave functions take the plane wave form in the $y$-direction.
Denoting the wave number by $k$, the quantum states are represented by
$|\alpha \rangle = |j,n,k,\sigma \rangle$
using the layer index $j$, the Landau level index $n$, and spin $\sigma = \pm$.
We do not include the Dirac point index because
the energy levels are independent of that index.
The degeneracy with respect to the Dirac points is included 
as an overall factor in $\sigma_{zz}$.
The eigen-energy is $E_{n\sigma} = E_n - \mu_B B \sigma$,
where $\mu_B$ is the Bohr magneton.
In the Zeeman energy term we have assumed that
the $g$-factor is equal to two 
because spin-orbit coupling is negligibly small.

\begin{figure}
   \begin{center}
    \includegraphics[width=0.8 \linewidth]{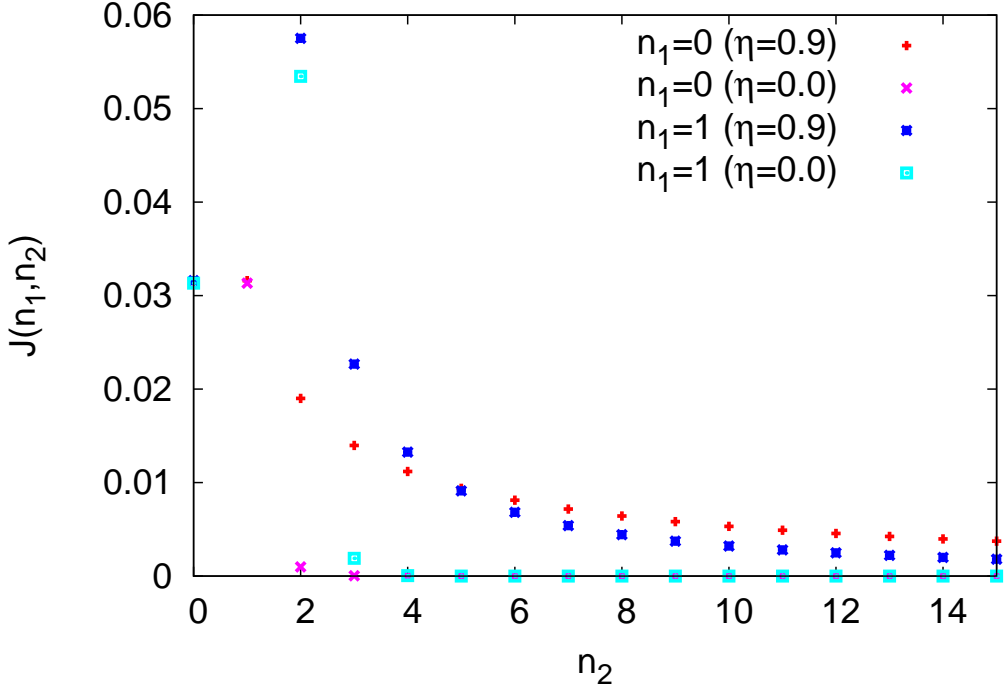}
   \end{center}
   \caption{ \label{fig_Jnn}
	(Color online)
	The current operator matrix element $J(n_1,n_2)$ as a function of 
	$n_2$ for $n_1=0$ and $n_1=1$ at $B=3{\rm T}$ ($c_y/\ell = 6.3 \times 10^{-2}$).
	Both of tilted and non-tilted Dirac cone cases are shown.
    }
 \end{figure}

Now we compute the matrix elements of the current operator.
Using the representation of the field operator 
in terms of the Landau level wave functions $\Phi_n(X)$,
defined by eq.\refeq{eq_Psi_n} with $\ell = 1/\sqrt{eB}$,
\be
\psi _{j,\sigma } \left( {x,y} \right) 
= \frac{1}{{\sqrt {\ell L_y } }}
\sum\limits_{n,k} {\exp \left( {iky} \right)
\Phi _{n} \left( {\frac{x}{\ell } + k\ell } \right) } c_{j,n,k,\sigma },
\ee
the current operator matrix elements are given by
\bea
\left\langle {j,n,k,\sigma } \right|{\hat J}_{c'} \left| {j + 1,n',k',\sigma '} \right\rangle  
&=& ia_z t_{c'} \delta _{k,k'} \delta _{\sigma ,\sigma '} \exp \left( {ikc_y } \right)
\nonumber \\
& & \times \frac{1}{\ell }\int {dx} \,\exp \left( {ic_y eBx} \right)
\Phi _{n}^\dag  \left( {\frac{x}{\ell } + k\ell } \right)
\Phi _{n'} \left( {\frac{x}{\ell } + k\ell } \right)
\nonumber \\
&=& 
\frac{ia_z t_{c'} \delta_{k,k'} \delta_{\sigma,\sigma'}}
{\sqrt {\left( {2 - \delta _{n,0} } \right)\left( {2 - \delta _{n',0} } \right)} }
\left[ {I_{n,n'}^{\left( {0,0} \right)}  
+ {\mathop{\rm sgn}} \left( {nn'} \right)I_{n,n'}^{\left( {1,1} \right)} } \right. 
\nonumber \\
& & \left. { - \eta {\mathop{\rm sgn}} \left( {n'} \right) I_{n,n'}^{\left( {0,1} \right)}  
- \eta {\mathop{\rm sgn}} 
\left( n \right)I_{n,n'}^{\left( {1,0} \right)} } \right],
\eea
where
\bea
I_{n,n'}^{\left( {s,s'} \right)} &=& \int {dX} 
\exp \left( {i\frac{{c_y }}{{\ell }}X} \right)
h_{\left| n \right| - s} \left( {X + \frac{\eta }{{1 - \eta ^2 }}{\mathop{\rm sgn}} 
\left( n \right)\sqrt {2\lambda ^3 \left| n \right|} } \right) \nonumber \\
& & \times h_{\left| {n'} \right| - s'} 
\left( {X + \frac{\eta }{{1 - \eta ^2 }}{\mathop{\rm sgn}} 
\left( {n'} \right)\sqrt {2\lambda ^3 \left| {n'} \right|} } \right),
\eea
with $h_n(X)$ being defined by eq.\refeq{eq_hn} in Appendix.
The function $h_n(X)$ are the eigen-functions of the harmonic oscillator,
${\cal H}_h=p^2/2+\lambda^2 X^2/2$.
For the calculation of the interlayer conductivity, we need 
$J(n_1,n_2)$ defined through
\be
|\left\langle {j,n_1,k,\sigma } \right|{\hat J}_{c'} \left| {j + 1,n_2,k,\sigma} 
\right\rangle |/(a_z t_{c'}) \equiv J(n_1,n_2).
\ee
Numerical evaluations of $J(n_1,n_2)$ are shown 
in Fig.\ref{fig_Jnn} for tilted ($\eta=0.9$) and non-tilted ($\eta=0$) cases
with $c_y = 9.3 \AA$.
One crucial observation is that $J(n_1,n_2) \neq 0$ for $n_1\neq n_2$.
By contrast, for $c_y=0$, $J(n_1,n_2) = 0$ if $n_1\neq n_2$.
For the tilted case, $J(n_1,n_2)$ gradually vanishes by increasing $n_2$.
For the non-tilted case, $J(n,n+1)$ is close to that for the tilted case.
However, $J(n,n+m) \ll J(n,n+1)$ for $m \geq 2$.

\begin{figure}
   \begin{center}
    \includegraphics[width=0.7 \linewidth]{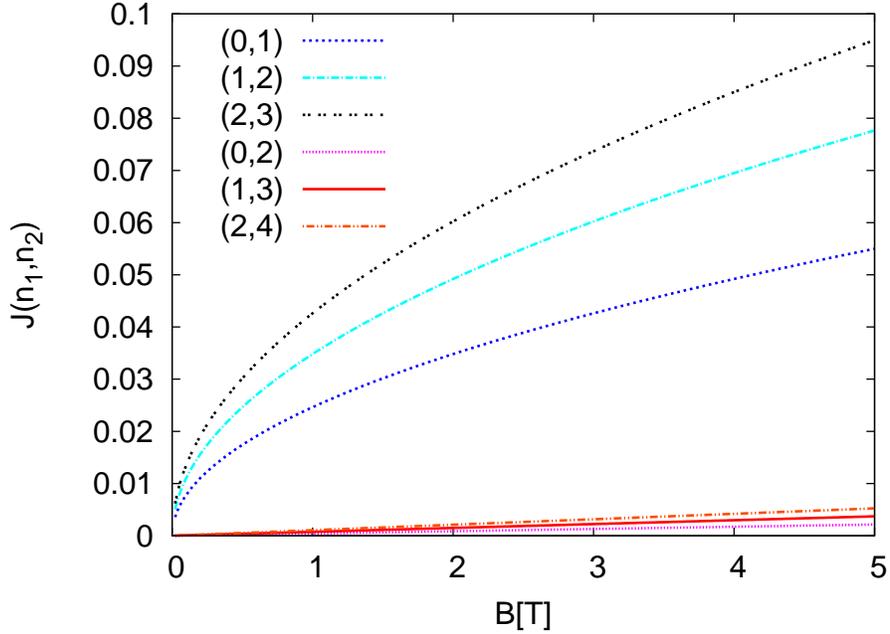}
   \end{center}
   \caption{ \label{fig_Jn1n2}
	(Color online)
	The magnetic field dependence of $J(n_1,n_2)$ for different $(n_1,n_2)$.
    }
 \end{figure}

For tilted cases we were unable to carry out the integral analytically.
For the non-tilted case, we rewrite the current operator matrix elements
in terms of a summation formula,
\be
J(n,n')  = \exp({ - \frac{1}{4}\left( {\frac{{c_y }}{\ell }} \right)^2 })
\left( {\frac{{c_y }}{\sqrt{2} \ell }} \right)^{\left| {n'} \right| 
- \left| n \right|} 
\left|
\sum\limits_{m = \max \left\{ {\left| n \right| 
- \left| {n'} \right|,0} \right\}}^{\left| n \right|} 
{\frac{{\sqrt {\left| n \right|!\left| {n'} \right|!} }}{{m!\left( {\left| n \right| - m} 
\right)!\left( {\left| {n'} \right| - \left| n \right| + m} \right)!}}
\left( { - \frac{1}{2}\left( {\frac{{c_y }}{\ell }} \right)^2 } \right)^m } \right|.
\label{eq_In1n2}
\ee
Note that $J(n,n)$ is analytically computed as 
$J(n,n)  = \exp \left( { -c_y^2/(4\ell^2) } \right)
| L_n \left( c_y^2/( 2\ell^2 ) \right) |$
with $L_n$ the $n$-th Laguerre polynomial.
For $n \neq n'$ we need to compute $J(n,n')$ numerically.
Figure \ref{fig_Jn1n2} shows the magnetic field dependence 
of the current operator matrix elements.
The matrix elements increase as we increase the magnetic field.
Note that the exponential factor in eq.(\ref{eq_In1n2}) is almost equal to one
because $c_y/\ell \ll 1$ for $B<500T$.
The parameter $c_y$ may depend on applied pressure.
However, the range of $c_y$ is much smaller than the range of 
the magnetic length $\ell$.
Since $J(n,n')$ is expressed as a function of $c_y/\ell$,
the change of $c_y$ is unimportant compared to the applied magnetic field change.

For the calculation of the interlayer conductivity, we need to include 
all matrix elements between different Landau levels:
If we include $n$ Landau levels, we need to compute $n(n-1)/2$ matrix elements.
Since the expression of the matrix element for the case of the tilted Dirac cone
contains the one-dimensional integral, the numerical calculation is heavy because
it turns out that we need $n \sim 100$ to observe the positive 
interlayer magnetoresistance region at low magnetic fields.
In order to focus on the effect of the non-vanishing matrix elements
we use the simpler expression eq.\refeq{eq_In1n2} for the non-tilted case
in the following.

\begin{figure}
   \begin{center}
    \includegraphics[width=0.7 \linewidth]{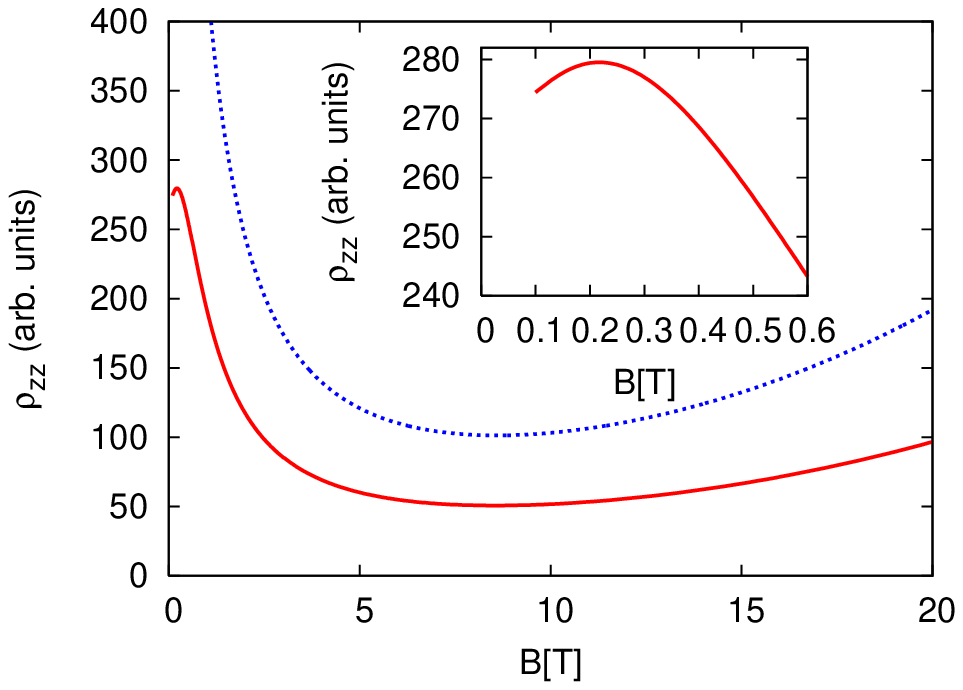}
   \end{center}
   \caption{ \label{fig_result}
	(Color online)
	Interlayer magnetoresistance for $C = 20{\rm K}{\rm T}^{-1/2}$ and $\Gamma=10$K 
	at zero temperature (solid line).
	The dotted line shows the case of $J(n,n') \rightarrow \delta_{n,n'}$.
	The inset shows the magnification around the peak.
    }
 \end{figure}

The interlayer magnetoresistance for $C=20 {\rm K}{\rm T}^{-1/2}$ and $\Gamma=10$K is
shown in Fig.\ref{fig_result} .
For $B > 10$T, there is a positive magnetoresistance region.
This positive magnetoresistance is associated with the Zeeman energy effect
as discussed by Osada.\cite{Osada08}
Our new finding is that the appearance of 
the positive magnetoresistance region in very low magnetic field.
For $B<0.2$T, there is another positive magnetoresistance region.
From the comparison with the case of $J(n,n') \rightarrow \delta_{n,n'}$
(dotted line in Fig.\ref{fig_result}),
it is clear that this positive magnetoresistance region
appears because of the presence of the non-vanishing current operator matrix 
elements between different Landau levels.

\begin{figure}
   \begin{center}
    \includegraphics[width=0.7 \linewidth]{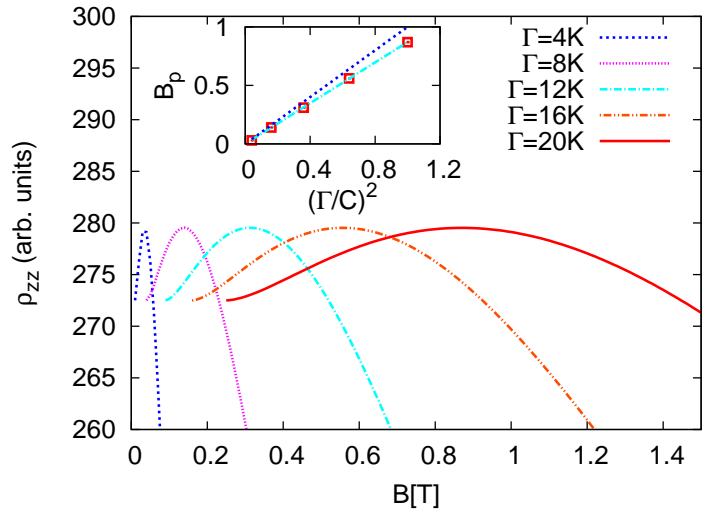}
   \end{center}
   \caption{ \label{fig_gammas}
	(Color online)
	Interlayer magnetoresistance for different values of $\Gamma=4,8,12,16,20$K
	with $C = 20{\rm K}{\rm T}^{-1/2}$.
	The inset shows $(\Gamma/C)^2$ dependence of the peak position, $B=B_p$.
	The dotted line shows $B_p = (\Gamma/C)^2$. (The dashed-dotted line
	is a guide to the eye.)
    }
 \end{figure}

In order to make clear how the peak position is determined,
we calculated the interlayer magnetoresistance for different values
of $\Gamma$ (Fig.\ref{fig_gammas}).
The peak appears at $B_p \simeq (\Gamma/C)^2$.
The appearance of the peak at $B=B_p$ is understood as follows.
For $B>B_p$, the zero-energy Landau level, $n=0$, is well 
separated from the other Landau levels because $E_1 - E_0 > \Gamma$.
Therefore, the non-vanishing $J(0,1)$ does not play an important role.
For $B<B_p$, there are large overlaps between Landau levels.
The Landau levels are mixed up upon interlayer hopping processes
because of non-vanishing $J(n,n')$.
The reason why $B_p$ is not exactly equal to $(\Gamma/C)^2$ 
is that the Landau levels are not equally separated in Dirac fermion systems.
That is, $E_{n+1} - E_n \neq E_1 - E_0$ for $n \neq 0$,
though the condition for the $n=0$ Landau level and the $n=1$ Landau level
plays the dominant role in the determination of the peak position.
The fact that $B_p < (\Gamma/C)^2$ is consistent with this picture.

\begin{figure}
   \begin{center}
    \includegraphics[width=0.7 \linewidth]{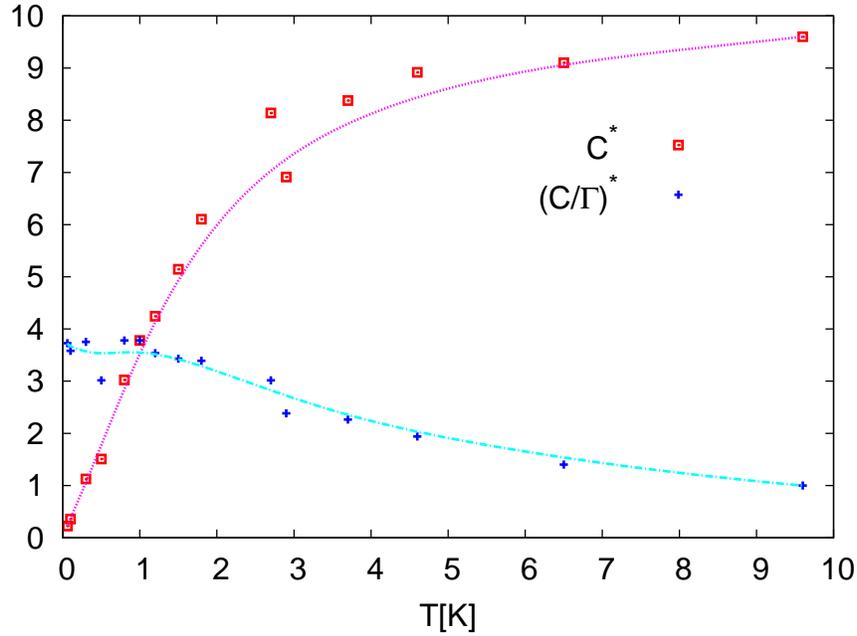}
   \end{center}
   \caption{ \label{fig_exp_analysis}
	(Color online)
	The tempeature dependence of $C^* \equiv k_B T/\sqrt{B_p}$
	and $(C/\Gamma)^* \equiv 1/\sqrt{B_p}$.
	The dotted line and the dashed-dotted line are guides to the eye.
    }
 \end{figure}

Now we analyze the experimental data in Ref.\citenum{Tajima09}.
The key fact is that the Landau level broadening 
arises from the temperature effect and the impurity scattering effect.
(Since the peak is observed for small magnetic fields,
we may neglect the Landau level broadening effect described by
the self-consistent Born approximation.\cite{Shon98})
At high temperatures, the Landau level broadening mainly 
arises from the temperature effect.
So we may neglect the impurity scattering effect.
The Landau level broadening factor $\Gamma$ is given by $\Gamma = k_B T$.
Therefore, if we apply the formula $B_p = (\Gamma/C)^2 = (k_B T/C)^2$
to the experimental data, we expect that $k_B T/\sqrt{B_p}$ approaches
a temperature independent constant value.
The constant value is approximately equal to $C$.
In order to estimate $C$,
we evaluated the peak position, $B_p$ 
from the experimental data (not shown),
and plotted $k_B T/\sqrt{B_p}$ in Fig.\ref{fig_exp_analysis}.
We see that this quantity approaches a temperature independent constant value.
From this plot, we found $C \sim 10{\rm K}{\rm T}^{-1/2}$.
By contrast, at low temperatures the Landau level broadening
arises from the temperature independent impurity scattering effect.
Therefore, we expect that $1/\sqrt{B_p}$ becomes temperature
independent at low temperatures.
In Fig.\ref{fig_exp_analysis}, $1/\sqrt{B_p}$ as a function
of temperature is shown.
We see that the data is almost temperature independent 
at low temperatures.
From this analysis we evaluated $\Gamma$ as $\Gamma \sim 3$K.

\section{Summary}
\label{sec_summary}
In this paper, we have derived a formula describing
the crossover from the positive to negative magnetoresistance
in the layered Dirac fermion system 
assuming a non-vertical interlayer tunneling.
Although the tunneling direction may depend on applied pressure,
the effect of the change of the tunneling direction 
is negligible compared to the change of the magnetic field
because the magnetic length is much longer than the change 
of the electron coordinate in the plane.
For a non-vertical interlayer tunneling,
there are inter-Landau level mixing upon interlayer tunneling.
This inter-Landau level mixing leads to 
the positive magnetoresistance at low magnetic fields.
(On the other hand, the positive magnetoresistance associated with
the Zeeman energy is observed at high magnetic fields
as pointed out in Ref.\citenum{Osada08}.)
The inter-Landau level mixing effect plays an important role
when the Landau level broadening factor is less than
the Landau level energy gap.
The crossover magnetic field is approximately given by
$B_p \simeq (\Gamma/C)^2$.
We have applied this formula to the experimental data.\cite{Tajima09}
The Dirac fermion spectrum parameter $C$ is evaluated to be 
$C \sim 10{\rm K}{\rm T}^{-1/2}$.
It would be interesting to compare the value of $C$ evaluated 
in this paper with that evaluated by other analyses.

\section*{Acknowledgment}
We would like to thank N. Tajima for helpful discussions and sending us
experimental data.
We also thank T. Osada, A. Kobayashi, and T. Himura for discussions.
This work was supported by the Grant-in-Aid for Scientific Research
from the Ministry of Education, Culture, Sports, Science and Technology (MEXT) 
of Japan, the Global COE Program 
"The Next Generation of Physics, Spun from Universality and Emergence," 
and Yukawa International Program for Quark-Hadron Sciences at YITP.


\appendix
\section{Landau level wave functions}
\label{sec_app}
In this appendix, we derive the Landau level wave functions
for tilted Dirac cone Hamiltonian \refeq{eq_DiracH}.
Under the magnetic field ${\bf B}=(0,0,B)$, 
taking the Landau gauge, ${\bf A}=(0,Bx,0)$, the Schr{\" o}dinger equation is
\be
v\left( {\begin{array}{*{20}c}
   {\eta \left( {{\hat k}_y  + eBx} \right)} & {{\hat k}_x  
- i\left( {{\hat k}_y  + eBx} \right)}  \\
   {{\hat k}_x  + i\left( {{\hat k}_y  + eBx} \right)} & 
{\eta \left( {{\hat k}_y  + eBx} \right)}  \\
\end{array}} \right)\psi \left( {x,y} \right) = E\psi \left( {x,y} \right).
\ee
Since the Hamiltonian has the translation symmetry in the $y$-direction,
we write the wave function $\psi (x,y)$ as
\be
\psi \left( {x,y} \right) = \frac{1}{\sqrt{L_y}}\exp \left( {iky} \right)\phi \left( x \right).
\label{eq_Sch1}
\ee
After substituting this form, we have the 
one-dimensional Schr{\" o}dinger equation,
\be
v\left( {\begin{array}{*{20}c}
   {\eta \left( {k + eBx} \right)} & { - i\frac{d}{{dx}} - i\left( {k + eBx} \right)}  \\
   { - i\frac{d}{{dx}} + i\left( {k + eBx} \right)} & {\eta \left( {k + eBx} \right)}  \\
\end{array}} \right)\phi \left( x \right) = E\phi \left( x \right).
\ee
Introducing $X=x/\ell + k\ell$ with $\ell=1/\sqrt{eB}$ the magnetic length,
\refeq{eq_Sch1} is rewritten as
\be
\left( {\hat M} + \eta X \right)\Phi \left( X \right) = \varepsilon \Phi \left( X \right),
\label{eq_Sch_a}
\ee
where $\Phi \left( X \right) = \sqrt \ell  \phi \left( x \right)$,
$\varepsilon  = E\ell /v$, and
\be
{\hat M} = \left( {\begin{array}{*{20}c}
   0 & { - i\frac{d}{{dX}} - iX}  \\
   { - i\frac{d}{{dX}} + iX} & 0  \\
\end{array}} \right).
\ee
Now we subtract the term, $\eta X \Phi (X)$, from the both hand sides of eq.\refeq{eq_Sch_a},
\be
{\hat M}\Phi \left( X \right) = \left( {\varepsilon  - \eta X} \right)\Phi \left( X \right).
\ee
This is the key step to solve the Schr{\" o}dinger equation.
Multiplying the operator ${\hat M}$ from the left hand side, and using
\be
{\hat M}\left( {\varepsilon  - \eta X} \right) 
= \left( {\varepsilon  - \eta X} \right){\hat M} + \left( {\begin{array}{*{20}c}
   0 & {i\eta }  \\
   {i\eta } & 0  \\
\end{array}} \right),
\ee
\be
{\hat M}^2  =  - \frac{{d^2 }}{{dX^2 }} + X^2  + \left( {\begin{array}{*{20}c}
   1 & 0  \\
   0 & { - 1}  \\
\end{array}} \right),
\ee
we obtain
\be
\left[ { - \frac{{d^2 }}{{dX^2 }} + \left( {1 - \eta ^2 } \right)
\left( {X + \frac{\eta }{{1 - \eta ^2 }}\varepsilon } \right)^2 } \right]
\Phi \left( X \right) = \left( {\begin{array}{*{20}c}
   {\frac{1}{{1 - \eta ^2 }}\varepsilon ^2  - 1} & {i\eta }  \\
   {i\eta } & {\frac{1}{{1 - \eta ^2 }}\varepsilon ^2  + 1}  \\
\end{array}} \right)\Phi \left( X \right).
\label{eq_Sch2}
\ee
The operator in the square bracket in the left hand side
has the form of the harmonic oscillator with the frequency 
$\lambda = \sqrt{1-\eta^2}$.
The eigen-values are $(2n+1)\lambda$ and
the eigen-functions are
\be
h_n \left( X' \right) = \frac{(-1)^n}{{2^{n/2} \pi ^{1/4} \sqrt {n!} }}\lambda ^{1/4} H_n 
\left( {\lambda ^{1/2} X'} \right)\exp \left( { - \frac{\lambda }{2}X'^2 } \right),
\label{eq_hn}
\ee
with $X'=X+\eta \epsilon/(1-\eta^2)$
and $H_n$ the $n$-th Hermite polynomial.
Making use of this result, we solve \refeq{eq_Sch2}.
The eigen-functions with the eigen-energy 
$\varepsilon = \pm 
\sqrt {2\lambda^3 n}$ are
\be
\frac{1}{\sqrt {2 + 2 \lambda } }
\left( {\begin{array}{*{20}c}
   {1 + \lambda }  \\
   { - i\eta }  \\
\end{array}} \right)h_{n-1} \left( X'  \right),
\ee
for $n \geq 1$ and
\be
\frac{1}{{\sqrt {2 + 2\lambda  } }}\left( {\begin{array}{*{20}c}
   {i\eta }  \\
   {1 + \lambda }  \\
\end{array}} \right)h_n \left( X'  \right).
\ee
Note that these degenerate eigen-functions are for eq.\refeq{eq_Sch2}.
In order to find the eigen-states of eq.\refeq{eq_Sch_a},
we take a linear combination of these eigen-functions.
After substituting the expression into eq.\refeq{eq_Sch_a}
and then after some algebra, we find the eigen-functions of eq.\refeq{eq_Sch_a},
\be
\Phi _n \left( X \right) = \frac{1}
{{\sqrt {2 - \delta _{n,0} } \sqrt {2\left( {1 + \lambda } \right)} }}
\left[ {\left( {\begin{array}{*{20}c}
   {i\eta }  \\
   {1 + \lambda }  \\

 \end{array} } \right)h_{\left| n \right|} \left( {X_n } \right) 
- i\operatorname{sgn} \left( n \right)\left( {\begin{array}{*{20}c}
   {1 + \lambda }  \\
   { - i\eta }  \\
 \end{array} } \right)h_{\left| n \right| - 1} \left( {X_n } \right)} \right],
\label{eq_Psi_n}
\ee
with 
\be
X_n  = X + \frac{\eta }
{{1 - \eta ^2 }}\operatorname{sgn} \left( n \right)\sqrt {2\lambda ^3 \left| n \right|},
\ee
and the eigen-energy, 
$\varepsilon _n  = \operatorname{sgn} \left( n \right)\sqrt {2\lambda ^3 \left| n \right|}$.

\bibliography{../../../references/tm_library2}


\end{document}